\documentclass[twocolumn,superscriptaddress,showpacs,prd,aps,amsmath,amssymb]{revtex4-1}
\usepackage{natbib}
\usepackage{graphicx,color}
\usepackage{amsmath,amssymb}
\usepackage{verbatim}
\usepackage{hyperref}
\usepackage{float}
\usepackage{wasysym}
\usepackage{amssymb,graphicx}
\usepackage{epsfig}
\usepackage{psfrag}
\usepackage{dsfont}
\usepackage{amsfonts}
\usepackage{mathrsfs}
\usepackage{times}

\begin{document}

\title{Are fast radio bursts the most likely electromagnetic
  counterpart \\ of neutron star mergers resulting in prompt collapse?}
\author{Vasileios Paschalidis} \affiliation{Departments of Astronomy
  and Physics, University of Arizona, Tucson, AZ 85719} \author{Milton
  Ruiz} \affiliation{Department of Physics, University of Illinois at
  Urbana-Champaign, Urbana, IL 61801}


\begin{abstract}
Inspiraling and merging binary neutron stars (BNSs) are important
sources of both gravitational waves and coincident electromagnetic
counterparts. If the BNS total mass is larger than a threshold value,
a black hole ensues promptly after merger. Through a statistical study
in conjunction with recent LIGO/Virgo constraints on the nuclear
equation of state, we estimate that up to $\sim 25\%$ of BNS mergers
may result in prompt collapse. Moreover, we find that most models of
the BNS mass function we study here predict that the majority of
prompt-collapse BNS mergers have $q\gtrsim 0.8$.  Prompt-collapse BNS
mergers with mass ratio $q \gtrsim 0.8$ may not be accompanied by
detectable kilonovae or short gamma-ray bursts, because they unbind a
negligible amount of mass and form negligibly small accretion disks
onto the remnant black hole. We call such BNS mergers
``orphan''. However, recent studies have found that
${10^{41-43}(B_p/10^{12}\rm G)^2 erg\, s^{-1}}$ electromagnetic
signals can be powered by magnetospheric interactions several
milliseconds prior to merger. Moreover, the energy stored in the
magnetosphere of an orphan BNS merger remnant will be radiated away in
${\mathcal O}(1\ \rm ms)$. Through simulations in full general
relativity of BNSs endowed with an initial dipole magnetosphere, we
find that the energy in the magnetosphere following black hole
formation is $E_B \sim 10^{39-41}(B_p/10^{12}\rm G)^2$ erg. Radiating
$\sim 1\%$ of $E_B$ in 1 ms, as has been found in previous studies,
matches the premerger magnetospheric luminosity. These magnetospheric
signals are not beamed, and their duration and power agrees with those
of non-repeating fast radio bursts (FRBs). These results combined with
our statistical study suggest that a non-repeating FRB {\it may} be
the most likely electromagnetic counterpart of prompt-collapse
BNSs. Detection of a non-repeating FRB coincident with gravitational
waves from a BNS merger could settle the extragalactic origin of a
fraction FRBs and could be used to place constraints on the nuclear
equation of state. FRBs can also initiate triggered searches for weak
signals in the LIGO/Virgo data.
\end{abstract}

\maketitle

\section{Introduction}

The LIGO and Virgo collaborations have already reported the direct
detection of gravitational waves (GWs) from the inspiral and merger of
a number of binary black
holes~\cite{FirstDirectGW,Abbott:2016nmj,Abbott:2017vtc,Abbott:2017oio,Abbott:2017gyy,LIGOScientific:2018jsj}
and one binary neutron star (BNS)~\cite{TheLIGOScientific:2017qsa}
(event GW170817), that was accompanied by multiple electromagnetic
(EM) counterparts~\cite{Monitor:2017mdv,GBM:2017lvd}.  The
consequences for astrophysics and fundamental physics from these
observations are far reaching, and it is a matter of time until the
detection of such compact binaries becomes routine.

Merging BNSs are not only important sources of GWs, but also sources
of coincident EM counterparts. These systems had long been suspected
as the progenitors of short gamma-ray bursts
(sGRBs)~\cite{EiLiPiSc,NaPaPi,
  Pacz86,Piran:2002kw,bergeretal05,Foxetal05,hjorthetal05,
  bloometal06,prs15,Ruiz:2016rai,Baiotti:2016qnr,Paschalidis:2016agf}. The
detection of the GW170817-counterpart
GRB170817A~\cite{Monitor:2017mdv} has provided the best evidence, yet,
that some sGRBs are powered by BNSs. BNSs are also sources of
kilonovae/macronovae~\cite{Lattimer1974ApJ...192L.145L,Li:1998bw}.
The association of kilonova AT 2017gfo/DLT17ck with
GW170817~\cite{GBM:2017lvd} has verified this expectation, too.

Merging BNSs may also be progenitors for {\em fast radio bursts}
(FRBs) -- a new class of radio transients lasting between a few to a
couple of tens of
milliseconds~\cite{Lorimer:2007qn,Thornton:2013iua}. So far 78 FRBs
have been
detected~\footnote{See~http://www.astronomy.swin.edu.au/pulsar/frbcat
  for an up-to-date-catalog of FRBs.}. The existence of two repeating
FRBs ``FRB121102''~\cite{Spitler:2016dmz} (which has also been detected
recently by CHIME~\cite{Josephy:2019ahz}) and ``FRB
180814.J0422+7''~\cite{Amiri:2019bjk} points to a non-catastrophic
origin as opposed to a collapse or merger, which suggests that there
may be at least two different classes of FRB progenitors.  Several
models have been proposed to explain FRBs including magnetar giant
flares, coherent radiation from magnetic braking at BNS merger,
blitzars (collapsing supramassive NSs), dark-matter induced
collapse of NSs, axion-miniclusters, newborn highly
magnetized NSs in supernova remnants, black hole--neutron
star batteries, charged black hole (BH) binaries, black hole current
sheets, black hole superradiance induced by
plasma~\cite{Popov:2007uv,Totani:2013lia,Falcke:2013xpa,Bramante:2014zca,
  Tkachev:2014dpa,Mingarelli:2015bpo,Liebling:2016orx,Marcote:2017wan,
  Zhang:2017ndi,Nicholl:2017slv,Conlon:2017hhi,Yamasaki:2017hdr,Piro:2017zec,Margalit:2018bje,Most:2018abt}.

Kilonovae from BNS mergers require dynamical ejection of matter during
merger and/or from an accretion disk by neutrino irradiation, see
e.g.~\cite{Metzger:2016pju} for a review. It is also widely accepted
that BNSs can generate sGRBs, if a jet is launched by the BH-disk
engine that forms following merger. Thus, in a scenario where a
negligibly small disk forms, and a negligible amount of mass escapes,
one may expect no sGRB and/or an undetectable kilonova from the BNS
event. We will refer to such ``kilonova-free'' and ``sGRB-free'' BNS
mergers that are detected in the GW spectrum as ``orphan''. However,
we stress the term orphan will be used to only mean that any potential
accompanying kilonova/sGRB is sub-threshold, and not that they do no
exist. Note also that there exist "orphan afterglows" of sGRBs, where
the gamma-rays are not detected (they are sub-threshold), but the
radio afterglow is detected (see,
e.g.,~\cite{Ghirlanda:2014fha}). But, are there any scenarios where
such orphan BNS mergers arise?

Numerical relativity simulations have shown that when the BNS total
mass ($M_{\rm tot}$) is greater than a threshold mass ($M_{\rm
  thres}$), a BH ensues in the first millisecond after merger. In this
prompt-collapse scenario a negligible amount of matter is ejected
dynamically~\cite{Hotokezaka2013} (see also~\cite{Dietrich:2016hky})
and a negligible amount of matter is available to form a
disk~\cite{STU1,STU2,lset08,Hotokezaka:2011dh,Hotokezaka2013}. Negligibly
small disks were also reported in~\cite{Ruiz:2017inq}, where it was
demonstrated that in prompt-collapse BNS mergers a jet cannot be
launched as opposed to the ``delayed'' collapse scenario which forms
massive disks~\cite{Ruiz:2016rai,Ruiz:2019ezy}. For illustration we
note that ejecta masses $\sim 0.025-0.05M_\odot$ are required to
explain the kilonova associated with
GW170817~\cite{Coulter:2017wya,Drout:2017ijr,Shappee:2017zly,Kasliwal:2017ngb,
  Tanaka:2017qxj,Arcavi:2017xiz,Pian:2017gtc,Smartt:2017fuw,Soares-Santos:2017lru,
  Nicholl:2017ahq,Cowperthwaite:2017dyu}, while typical ejecta from
equal-mass, prompt-collapse BNS mergers are ${\mathcal
  O}(10^{-4}M_\odot)$ or
less~\cite{Hotokezaka2013,Radice:2017lry}~\footnote{Such small ejecta
  masses constitute $\sim 0.01\%$ of the total rest-mass and it is not
  clear that numerical relativity simulations have achieved such high
  levels of accuracy, yet.}, and disk masses ${\mathcal
  O}(10^{-3}M_\odot)$~\cite{Hotokezaka:2011dh}. According
to~\cite{Metzger:2011bv} ejecta masses ${\mathcal O}(10^{-3}M_\odot)$
or greater are required for detectable kilonovae at the depth and
cadence of the normal LSST survey with current or planned
telescopes. Therefore, prompt-collapse BNS mergers may appear orphan
unless they take place nearby. This raises the main question that we
focus on in this paper: what is the most likely electromagnetic
counterpart of orphan prompt-collapse BNS mergers?

First, we point out that if the binary mass ratio $q$ (defined here to
be less than unity) is smaller than 0.8, then both appreciable
matter may become unbound and a sizable disk onto the remnant BH may
form~\cite{Rezzolla:2010fd,Hotokezaka2013,Dietrich:2016hky}. This is
because for substantially asymmetric BNSs the lighter companion is
tidally disrupted before merger, in contrast to near equal-mass
binaries. Thus, sufficiently asymmetric, prompt-collapse BNS mergers
may power both sGRBs and kilonovae.

In this work we perform a statistical study to assess the
astrophysical relevance of prompt-collapse BNSs, and the likelihood of
orphan BNS mergers. In particular, we compute the $M_{\rm tot}$ and
$q$ distribution of BNSs using the Galactic NS mass function and
population synthesis models in conjunction with GW170817 constraints
on the nuclear equation of state (EOS). We estimate that up to $\sim
25\%$ of all BNSs may result in prompt collapse. We also find that
most models of the BNS mass function we treat predict that the
majority of prompt-collapse BNSs have $q\gtrsim 0.8$. Furthermore, the
larger $M_{\rm thres}$ is, the more skewed toward $q=1$ the
distribution of binaries with $M_{\rm tot} > M_{\rm thres}$
becomes. Thus, most prompt-collapse BNSs may appear orphan.
But, does this imply no detectable EM counterparts from such mergers?

Recent work found that interactions in compact binary
magnetospheres~\cite{Palenzuela:2013hu,Palenzuela:2013kra,Paschalidis:2013jsa,Ponce:2014sza,Ponce:2014hha}
(see
also~\cite{Hansen:2000am,McWilliams:2011zi,2012ApJ...755...80P,Lai:2012qe}
for related discussions) can power $\sim {10^{41-43}(B_p/10^{12}\rm
  G)^2 erg\, s^{-1}}$ EM signals several milliseconds prior to
merger. Here $B_p$ is the magnetic field strength at the pole of the
NS.  Moreover, following BH formation there is a significant
amount of energy stored in the magnetosphere of the remnant. Studies
of magnetospheres of stars collapsing to
BHs~\cite{Baumgarte:2002vu,Lehner:2011aa,Dionysopoulou2013} have shown
that a fraction $\epsilon \gtrsim 1\%$~\footnote{The fraction is $20\%$ for
  electrovacuum~\cite{Baumgarte:2002vu}} of the total energy stored in
a force-free magnetosphere is radiated away on a collapse timescale
$\tau_{\rm FRB}$. This timescale is $\mathcal{O}(1\ \rm ms)$ for a
NS. For a magnetic dipole in flat spacetime the total magnetic energy
in the magnetosphere is
\begin{eqnarray}\label{EB}
  E_{\rm B}  & \sim & \int_R^{\infty}\int_0^\pi \frac{B^2}{8\pi} \left(\frac{R}{r}\right)^6 \frac{5+3\cos(2\theta)}{8} 2\pi r^2 \sin\theta dr d\theta  \nonumber \\
  & \sim & \frac{1}{12}B^2R^3 \sim 10^{41} B_{12}^2 R_{10}^3\ \rm erg,
\end{eqnarray}
implying an outgoing EM luminosity of
\begin{equation}\label{LFRB}
L_{\rm FRB}
\sim 10^{42} \epsilon_{0.01}B_{12}^2 R_{10}^3\tau_{\rm  FRB,1}^{-1}\ \rm erg\ s^{-1}.
\end{equation}
Here, $B_{12}=B_p/10^{12}\ \rm G$, $R_{10}$ the stellar radius in
units of 10 km, $\epsilon_{0.01}$ the efficiency $\epsilon$ normalized
to 0.01, and $\tau_{\rm FRB,1}$ the emission time in units of 1
ms. Note that for a rotating collapsing star the efficiency is
$\epsilon\simeq 18\%$~\cite{Lehner:2011aa}, but here and throughout we
adopt the lower value $\sim 1\%$ as a lower bound.  This outgoing
luminosity in Eq.~\eqref{LFRB} matches the premerger magnetospheric
luminosity.  Moreover, the power and duration of these magnetospheric
signals match those of observed FRBs~\cite{Mingarelli:2015bpo}. Thus,
BNSs are candidates for non-repeating, FRBs, as has also been
suggested in~\cite{Totani:2013lia}.

Note that when two NSs merge and promptly collapse to a BH, the total
energy stored in the magnetosphere is anticipated to be of the same order
of magnitude as in Eq.~\eqref{LFRB}, because there is little time
available to amplify the surface magnetic field through hydromagnetic
instabilities as in a delayed collapse
scenario~\cite{Kiuchi:2015sga}. However, compression due to the
collision can amplify the magnetic field because of magnetic flux
freezing. On the other hand, a large amount of the energy will quickly
fall into the remnant BH. Thus, a detailed numerical relativity study
of prompt-collapse BNS mergers is necessary to assess the post-merger
magnetospheric energy of BNSs resulting in prompt collapse.

To confirm the expectation from Eq.~\eqref{LFRB}, we perform fully
general relativistic, ideal magnetohydrodynamics simulations of
prompt-collapse BNS mergers. Following BH formation we compute the energy
stored in the magnetosphere. Assuming a $1\%$ radiation efficiency and
a millisecond emission time, we estimate an outgoing burst with
luminosity $L_{\rm EM} \sim 10^{40-42} (\epsilon/0.01)(B/10^{12}\rm
G)^2$ erg/s, which at the edge of the LIGO BNS range translates to
flux densities of 0.1 to 30 $\rm Jy$ -- observable by existing radio
telescopes. Thus, our simulations provide support to the idea that the
collapse in prompt-collapse BNSs is a promising FRB counterpart to the
GWs, i.e., the FRB would not be only precursor, but continue also
after the peak GW amplitude.

To sum, BNS mergers are promising candidates for non-repeating FRBs,
and such FRBs {\it may} be the most promising EM counterpart of orphan
BNS mergers. The outgoing magnetospheric burst is rather
isotropic~\cite{Lehner:2011aa,Palenzuela:2013kra,Paschalidis:2013jsa},
in contrast to a sGRB which is beamed, making the detection of such
FRB signatures largely independent of the binary
orientation. Detection of an FRB can trigger searches in LIGO/Virgo
data. The discovery of coincident GWs with an FRB may settle the
extragalactic origin of a fraction of FRBs. Moreover, detection of an
FRB from an orphan BNS merger could provide strong evidence that the
merger resulted in prompt collapse to a BH, and could place
constraints on the nuclear EOS, see
e.g.~\cite{Bauswein:2013jpa}\footnote{Note that if one knows the
  collapse time from merger constraints on the EOS can be placed using
  sGRBs~\cite{Lasky:2013yaa}, too.}.  Note that without an
electromagnetic counterpart, a prompt-collapse BNS system might also
be interpreted as a low-mass binary black hole or other dark binary
compact object, because finite size effects become significant late in
the binary inspiral, where current gravitational wave detectors are
not as sensitive.  Thus, to discern a binary black hole from a
prompt-collapse BNS merger, the lack of a kilonova and sGRB is only a
necessary ingredient. The FRB would be important to solidify that
matter was present in the event and hence endorse information coming
from GWs on finite size effects. By contrast a near equal-mass binary
black hole-neutron star (BHNS) is likely to form accretion disks and
eject matter more than $10^{-3}M_\odot$, and hence power detectable
kilonovae. In particular, using the updated formula
of~\cite{Foucart:2018rjc} for the amount of mass outside the BH in a
BHNS merger, we find that for an equal-mass BHNS merger, adopting a
range of NS radii favored by GW170817~\cite{Abbott:2018exr}), i.e.,
compactness values $C_{\rm NS}\sim 0.165-0.205$, and BH spins
$\chi=0-0.93$ more than 90\% of the $C_{\rm NS}-\chi$ parameter space
results in mergers with mass outside the BH exceeding
$10^{-2.5}M_\odot$. Given that recent
work~\cite{Kasen:2014toa,Fernandez:2015use,Siegel:2017nub} has shown
that several tens of per cent of the mass outside the BH becomes
unbound due to viscous/magnetic/neutrino processes, the above imply
that near equal mass BHNSs most likely power observable kilonovae, and
possibly also short gamma-ray bursts. Thus, a prompt collapse BNS
merger can in principle be distinguished from a BHNS merger, after the
compact binary parameters have been inferred from the GW
observations. In addition, a prompt collapse merger is distinguishable
from a delayed collapse or no-collapse BNS merger, since numerical
simulations of such mergers show that delayed collapse or no-collapse
is associated with dynamical ejecta masses that are $>0.001M_\odot$
and disk masses of a few \% (see
e.g.,~\cite{Hotokezaka2013,Radice:2017lry,East:2019lbk}). Note also,
that the consensus in the community is that GW170817 was a delayed
collapse
merger~\cite{Ruiz:2017due,Rezzolla:2017aly,Margalit2017,Shibata:2017xdx,Bauswein:2017vtn}.
Hence, BHNS, delayed collapse and no collapse BNS mergers are all
anticipated to have detectable kilonovae, and thus are in all
likelihood distinguishable in this respect from prompt collapse BNS
mergers.


The remainder of the paper is organized as follows. In
Sec.~\ref{sec:Prob} prompt-collapse BNS mergers are motivated through
a study of the BNS $M_{\rm tot}$ and $q$ distribution. A description
of our simulations and results are presented in
Sec.~\ref{sec:sims}. Our conclusions are provided in
Sec.~\ref{sec:conclusion}. Geometrized units ($G=c=1$) are adopted
throughout, unless otherwise specified.

\section{Probability Estimates for BNS mergers}

\begin{figure*}
 \centering
 \includegraphics[width=0.48\textwidth]{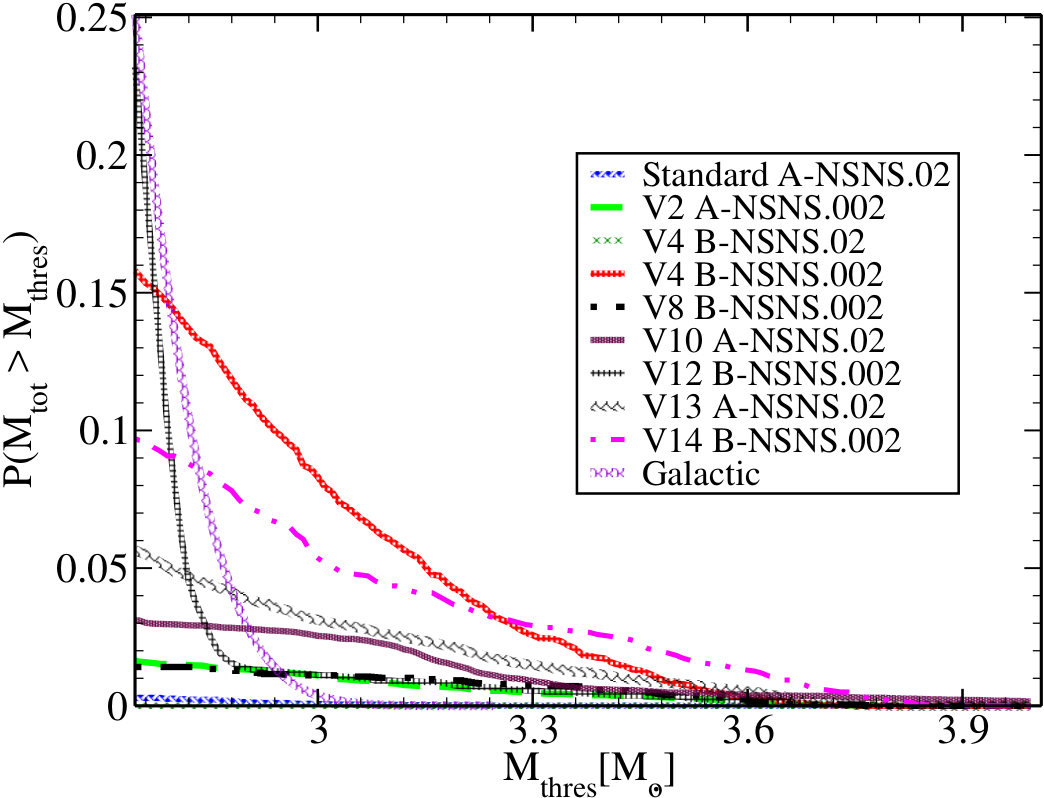}
 \hspace{0.5cm}\includegraphics[width=0.48\textwidth]{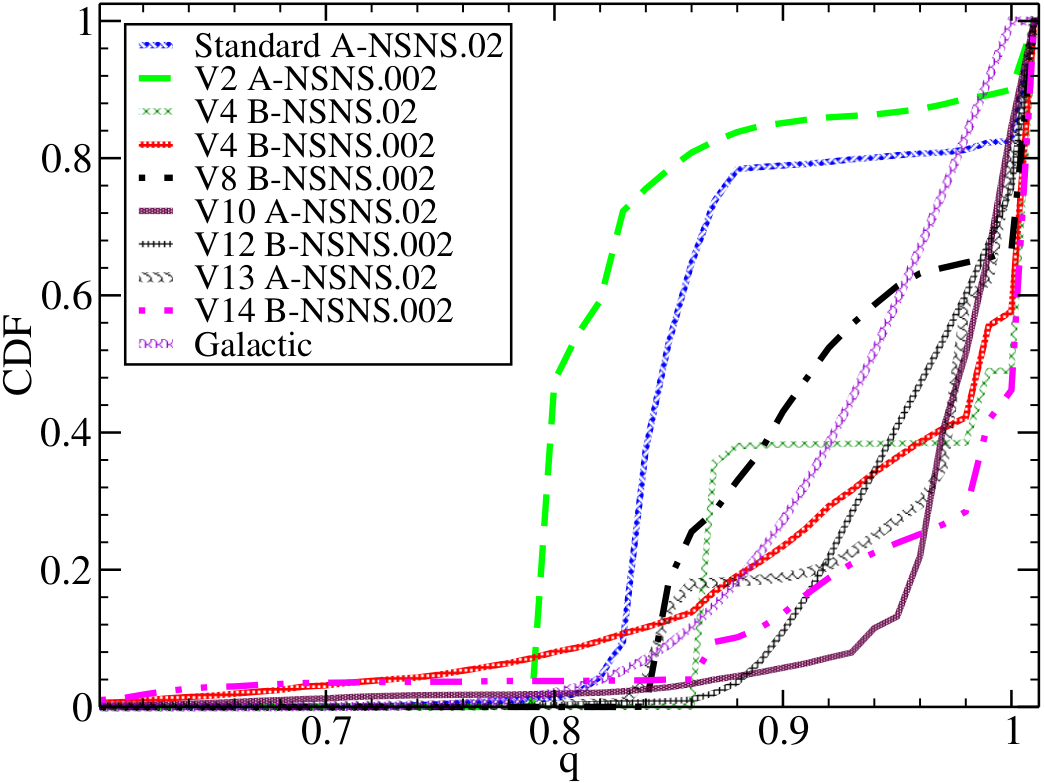} 
 \caption{Left: Probability for $M_{\rm tot}>M_{\rm thres}$, where
   $M_{\rm tot}$ is the binary ADM mass, if the binary components were
   infinitely separated. The curves labeled ``\#\#-\#NSNS.\#\#\#''
   correspond to population synthesis calculations, and the curve
   labeled ``Galactic'' corresponds to the mass distribution of
   Eq.~\eqref{NSmassDistrib}. Right: the cumulative distribution
   function (CDF) for the mass ratio that corresponds to the same
   models shown on the left.
  \label{fig:NSNS_distrib}}
\end{figure*}

\label{sec:Prob}

To assess whether prompt-collapse BNS mergers are astrophysically
relevant, and in particular whether {\it orphan} BNS mergers are
likely, we need to know the value of $M_{\rm thres}$, and the BNS
$M_{\rm tot}$ and $q$ distribution. We address these topics in this
section.

\subsection{Constraints on the threshold mass for prompt collapse}

While $M_{\rm thres}$ has been found to be independent of the mass
ratio~\cite{Shibata:2006nm}, it is sensitive to the nuclear EOS
\cite{STU1,STU2,Shibata:2006nm,Bauswein:2013jpa,Bauswein:2017aur},
which is not very well constrained, yet. A number of studies have
recently placed constraints on the nuclear EOS using the observation
of GW170817 (see, e.g.,~\cite{Paschalidis:2017qmb} and references
therein as well as~\cite{Raithel:2019uzi,Gandolfi:2019zpj} for
reviews). Here we focus on works that set constraints on the
Tolman-Oppenheimer-Volkoff (TOV) limit ($M_{\rm TOV}$), i.e., the
maximum mass supported by a non-rotating NS. In
particular,~\cite{Ruiz:2017due,Rezzolla:2017aly,Margalit2017,Shibata:2017xdx}
following different approaches concluded that GW170817 sets an upper
bound $M_{\rm TOV} \lesssim 2.2 M_\odot$ (\cite{Margalit2017} argues
for $M_{\rm TOV} \lesssim 2.17 M_\odot$ at 90\% confidence).  We now
use the upper bound on $M_{\rm TOV}$ to obtain a reasonable range for
$M_{\rm thres}$.

In~\cite{Bauswein:2013jpa} $M_{\rm thres}$ was computed for a number
of realistic, finite temperature EOSs, and was found that $M_{\rm
  thres} \in [2.95,3.85]M_\odot$. However, if we demand that the EOS
respect $M_{\rm TOV} \lesssim 2.2 M_\odot$, then the range shrinks to
$M_{\rm thres} \in [2.95,3.25]M_\odot$ for the EOSs considered
in~\cite{Bauswein:2013jpa}.

In addition,~\cite{Bauswein:2013jpa} derived the following
EOS-independent relation that expresses $M_{\rm thres}$ in terms of
$M_{\rm TOV}$~\footnote{An expression $M_{\rm thres}=kM_{\rm TOV}$
  was first proposed by~\cite{STU1}.}
\begin{equation}
  \label{MthresvsMtov}
M_{\rm thres} = (a\,C^*_{1.6} + b)M_{\rm TOV},
\end{equation}
where $a=-3.606$, $b=2.380$, and $C^*_{1.6} = M_{\rm TOV}/R^*_{1.6}$,
with $R^*_{1.6}$ the radius of a $1.6M_\odot$ NS for a given EOS. We
note that $M_{\rm thres}$ here is defined as the Arnowitt-Deser-Misner
(ADM) mass of the binary, if the binary companions were infinitely
separated. We can use Eq.~\eqref{MthresvsMtov} in conjunction with the
EOSs that are favored by GW170817~\cite{Abbott:2018exr} to explore how
small the lower bound on $M_{\rm thres}$ can become. We investigated
the masses and radii of cold nuclear EOSs listed
in~\cite{Read:2008iy}. Among the EOSs that respect $1.97M_\odot
\lesssim M_{\rm TOV} \lesssim 2.2M_\odot$~\footnote{The lower limit
  comes from the $2M_\odot$ pulsar
  observations~\cite{Demorest2010,Antoniadis2013}.}, and the
mass-radius constraints of~\cite{Abbott:2018exr}, the EOS
WFF1~\cite{WFF1} yields a smallest value for $M_{\rm thres}$ through
Eq.~\eqref{MthresvsMtov}; namely, $M_{\rm thres}\simeq
2.75M_\odot$. This is not unexpected because Eq.~\eqref{MthresvsMtov}
predicts that the softer the EOS (larger $C^*_{1.6}$) and the smaller
$M_{\rm TOV}$ are, the smaller $M_{\rm tresh}$ becomes. WFF1 is among
the softest EOSs with $M_{\rm TOV}\sim 2.0 M_\odot$.  Thus, in this
work we adopt $[2.75,3.25]M_\odot$ as a reasonable range for $M_{\rm
  thres}$ respecting current
constraints on the nuclear EOS.

\subsection{Binary neutron star total mass and mass-ratio distributions}

The NS mass function for Galactic BNSs has been modeled
in~\cite{Kizilta2013,Ozel:2016oaf}. As in~\cite{Yang:2017xlf}, in our
analysis below we use the Gaussian mass function
of~\cite{Ozel:2016oaf}, because it is simpler to work with and because
the skewed Gaussian of~\cite{Kizilta2013} is consistent with 0
skewness parameter, and hence agrees very well with the distribution
of~\cite{Ozel:2016oaf}. In~\cite{Ozel:2016oaf} the probability
distribution function of NS masses ($M_{\rm NS}$) in Galactic BNSs is
modeled as
\begin{equation}
  \label{NSmassDistrib}
P(M_{\rm NS};M_0,\sigma)=\frac{1}{2\pi \sigma^2}\exp\left[-\frac{(M_{\rm NS}-M_0)^2}{2\sigma^2}\right]
\end{equation}
with $M_0=1.33M_\odot$, and $\sigma=0.09M_\odot$. Assuming that the
masses of the two NSs in a BNS are independent random variables, we
can use Eq.~\eqref{NSmassDistrib} to derive the distribution of the
BNS $M_{\rm tot}$ and that of $q$. The $M_{\rm tot}$ distribution is
again given by Eq.~\eqref{NSmassDistrib}, but with $M_0=2.66M_\odot$,
$\sigma=0.09\times \sqrt{2}M_\odot$, and $M_{\rm NS}$ replaced with
$M_{\rm tot}$. Using the $M_{\rm tot}$ distribution we can compute the
probability that $M_{\rm tot}$ is greater than a certain value. In the
left panel of Fig.~\ref{fig:NSNS_distrib} this is shown by the curve
labeled ``Galactic'', which demonstrates that if $M_{\rm
  thres}=2.75M_\odot$, as in the WFF1 EOS, then $\sim 25\%$ of all
binaries result in prompt collapse. However, if $M_{\rm
  thres}=3.25M_\odot$ (the upper value in the range we discussed in
the previous subsection), then the Galactic NS mass function predicts
that there are practically no BNSs resulting in prompt collapse. If we
use $M_{\rm thres}\simeq 2.8$~\cite{Hotokezaka:2011dh}, which
corresponds to the SLy~\cite{Douchin:2001sv} and APR4~\cite{Akmal98}
EOSs, also favored by GW170817~\cite{Abbott:2018exr}, then the
Galactic NS mass function predicts that $\sim 13.5\%$ of all BNSs
result in prompt collapse.

The Galactic mass function may not be representative of all
BNSs. Thus, we also use results from population synthesis
studies~\cite{Dominik:2012kk}. In the left panel of
Fig.~\ref{fig:NSNS_distrib} we show the probability that $M_{\rm
  tot}>M_{\rm thres}$ for one of the standard models
of~\cite{Dominik:2012kk} labeled ``Standard'', and several variations
of the standard models labeled ``\#\#-\#NSNS.\#\#\#''
(see~\cite{Dominik:2012kk,SyntheticUniverse} for the labeling and what
parameters are varied). The conclusion from the plot is that there are
realizations with a wide tail at large $M_{\rm tot}$, for which a
significant fraction of BNSs result in prompt collapse (even for
$M_{\rm thres} = 3.25M_\odot$). However, there exist realizations for
which there are practically no BNSs with $M_{\rm tot}>M_{\rm thres}$
(even for $M_{\rm thres} = 2.75M_\odot$). But, the fact that GW170817
favors softer EOSs, makes prompt-collapse BNS mergers potentially
observationally relevant.

Next we address whether any orphan prompt-collapse mergers are
expected. As mentioned above, we anticipate that prompt-collapse BNS
mergers will eject appreciable matter and form disks for $q <
0.8$. Using Eq.~\eqref{NSmassDistrib} for the Galactic NS mass
distribution in BNSs we can compute the $q$ distribution of BNSs. In
the right panel of Fig.~\ref{fig:NSNS_distrib} we show the cumulative
distribution of $q$ for Milky-way like BNSs labeled
``Galactic''. Thus, for the Galactic mass function more than $\sim
80\%$ of BNSs have $q>0.9$. We have also checked that this result
holds even when restricting to binaries with $M_{\rm tot}$ greater
than $M_{\rm thres} \in [2.75M_\odot, 3.25M_\odot]$. Moreover, we find
that for larger $M_{\rm thres}$, the $q$ distribution of $M_{\rm
  tot}>M_{\rm thres}$ binaries is skewed even more toward $q=1$. This
result is explained as follows: the number of very high mass NSs is
very low, and achieving $M_{\rm tot}$ more than $\sim 3.00M_\odot$
requires $q\sim1$ binaries.

The $q$ distribution from select population synthesis models is also
shown in the right panel of Fig.~\ref{fig:NSNS_distrib}. It is clear
that $q \gtrsim 0.8$ in most cases, and there exist realizations where
more than $\sim 90\%$ of BNSs have $q> 0.95$. We have also checked
that these results hold, even when restricting to binaries with
$M_{\rm tot} > M_{\rm thres}$. As in the Galactic case, we find in the
population synthesis results, too, that the larger $M_{\rm thres}$ is,
the more symmetric binaries with $M_{\rm tot}>M_{\rm thres}$ become.
In particular of all 60 variations of populations synthesis models
available in~\cite{SyntheticUniverse}, we find that for $M_{\rm
  tot}>M_{\rm thres}$ only 17, 15 and 3 variations have 20\% or more
binaries with $q<0.8$, for $M_{\rm thres}=2.75, 2.95$, and
$3.25M_\odot$, respectively.

These results and the discussion in the previous section suggest that
the majority of prompt-collapse BNS mergers are likely to appear
orphan, and hence their most promising EM counterpart likely will
arise by magnetospheric effects, and may be a non-repeating FRB.

%
%
\begin{center}
  \begin{table*}[t]
    \caption{Summary of main results. Here $E_B$ is the magnetic
      energy stored in the magnetosphere as measured by observers
      comoving with the plasma $t\sim 200M$ following BH
      formation. $L_{\rm FRB}$ is the estimated luminosity produced by
      the ejection of $0.8\%$ of the magnetic energy stored in the
      magnetosphere in $\tau_{\rm FRB}=1\rm ms$. $S_{\nu}$ is the flux
      density at the detector in units of $\rm Jy$ assuming a nominal
      observing frequency of 1 GHz and that the source is located at
      the edge of the LIGO BNS range, i.e., 200Mpc. Units are assigned
      by setting the polytropic constant $k=262.7\rm km^2$.
      \label{table:results}}
    \begin{tabular}{cccc}
      \hline\hline 
          {Case Model} & $E_B/B_{12}^2$ $\rm [ erg]$ & $L_{\rm FRB}/(B_{12}^2 \tau_{\rm FRB,1}^{-1})$ $\rm [ erg\,s^{-1}]$ & $S_{\nu}$[Jy] at $\nu=1\ \rm GHz$ \\ 
          \hline
          P-Prompt-1  & $10^{40.9}$ & $10^{41.8}$ & 13.2 \\
          P-Prompt-2  & $10^{38.9}$ & $10^{39.8}$ & 0.13  \\
          P-Prompt-3  & $10^{41.3}$ & $10^{42.2}$ & 33.1 \\
\hline\hline 
\end{tabular}
 \end{table*}
\end{center}
%

\section{Simulations and results}
\label{sec:sims}

We performed fully general relativistic, ideal magnetohydrodynamic
simulations of BNSs endowed with an initial dipole magnetosphere to
assess whether prompt-collapse BNSs have enough energy stored in the
remnant magnetosphere to power an FRB. We adopt the code
of~\cite{APPENDIXPAPER,UIUC_PAPER1,Farris:2012ux}.  Our evolution
methods and grid set up are the same as those described
in~\cite{Ruiz:2017inq}. The initial data we adopt are publicly
available, have been generated with the {\tt LORENE}
library~\cite{lorene} and correspond to cases P-Prompt-1, P-Prompt-2,
and P-Prompt-3 of~\cite{Ruiz:2017inq}. These are $\Gamma=2$
polytropic~\footnote{The EOS is $P=k\rho_0^\Gamma$ with $P$ the
  pressure, $\rho_0$ the rest-mass density, and $k$, $\Gamma$ the
  polytropic constant and exponent, respectively}, irrotational BNS
initial data. We seed an initial dipole magnetic field in each NS by
use of Eq.~(2)~of~\cite{Paschalidis:2013jsa}. The resulting magnetic
field configuration is the same as in~\cite{Ruiz:2017inq}, but we set
the initial polar magnetic field (as measured by comoving observers)
to $B_{p}=10^{12}$ G.  This initial magnetic field is dynamically
unimportant, thus our simulations scale with $B_p$. In our results
below we show the scaling with $B_{12}=B_p/10^{12}\ \rm G$. To mimic
the force-free conditions in NS magnetospheres we adopt the
method we developed in~\cite{prs15} where at t=0 we impose a low but
variable density atmosphere with a universal plasma parameter beta
less than unity. The value of the plasma beta is 0.01, and this
captures one key aspect of force-free electrodynamics, i.e., magnetic
field pressure dominance. As explained in~\cite{prs15} our code can
handle such values of plasma parameter beta~\footnote{Note that as
  shown in~\cite{McKinney:2006sc,Paschalidis:2013gma} force-free
  electrodynamics is a subset of ideal magnetohydrodynamics. Thus in
  principle a perfect ideal magnetohydrodynamics (MHD) code can also
  model force-free environments. However, limitations in current
  numerical schemes of ideal MHD do not allow the evolution of highly
  magnetized matter. But, they do allow the evolution of moderately
  magnetized matter.}.

The basic dynamics of these systems has been described
in~\cite{Ruiz:2017inq} where it was shown that these systems form
negligibly small disks onto the remnant BH and no jets are
launched. We terminate our simulations when the electromagnetic
energy outside the remnant BH has settled. We compute the energy
stored in the magnetosphere as measured by comoving observers as in
Eq. (9) of~\cite{Ruiz:2017inq}. At any given time we compute the
magnetospheric energy ($E_B$) only below a certain rest-mass density
which we set to $10^{-4}$ of the maximum rest-mass density on the grid
at that time. For case P-Prompt-3 we also changed this value to
$10^{-5}$ of the maximum density to test if this choice makes a
difference. We call this case P-Prompt-3$^*$. We list the measured
energy in the magnetosphere outside the BH after it has settled
in~Table~\ref{table:results}. As is clear from the table the energy
matches well the order-of-magnitude predictions of Eq.~\eqref{EB}.

\begin{figure}
 \centering \includegraphics[width=0.48\textwidth]{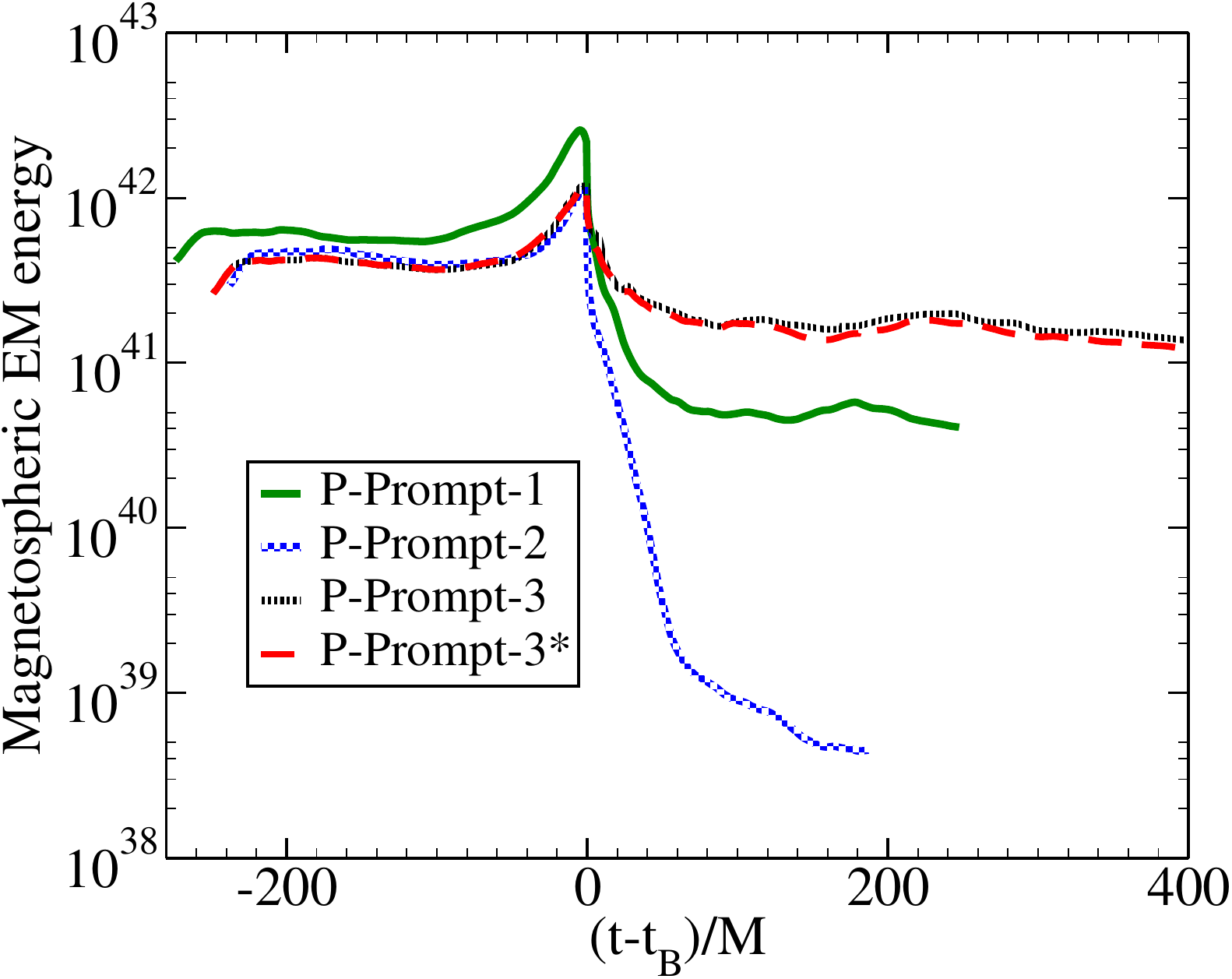}
 \caption{Time evolution of electromagnetic energy in the
   magnetosphere for the 4 cases studied in this work. The time axis
   is shifted with respect to the time of BH formation $t_B$ and is
   normalized to the ADM mass of the system. Physical units are
   assigned by setting the polytropic constant $k=262.7\rm km^2$.
  \label{fig:EMenergy}}
\end{figure}

In Fig.~\ref{fig:EMenergy} we show the time evolution of the EM energy
in the magnetosphere for each case we considered. The plot exhibits
that after an initial settling of the magnetosphere, the
magnetospheric energy is approximately constant until merger, at which
point it increases by about a factor of 2 by the collision, and
subsequently decays, as part of the magnetosphere flows into the
remnant BH. Cases P-Prompt-3 and P-Prompt-3* demonstrate that changing
the cut-off density by an order of magnitude for the computation of
the electromagnetic energy in the magnetosphere has an effect that is
less than 5\% up to $200M$ following BH formation when the EM has
already approximately settled.

It is not clear where the difference in the electromagnetic energy in
the magnetospheres in the 3 cases we study is coming from. However,
the different configurations undergo collapse in different ways,
because P-prompt-2 is much more massive than the other two cases, and
P-prompt-1 is asymmetric. It is likely that the more ``violent''
collapse of case P-prompt-2 drags a larger part of the magnetosphere
through the horizon leaving less electromagnetic energy exterior to
the BH. This suggests that the ``promptness'' of the collapse may
determine the amount of energy in the magnetosphere. More detailed
studies are necessary to solidify this explanation, and these will be
the subject of future work.

To estimate the outgoing EM luminosity that is expected to be produced
by the ``release'' of the magnetosphere, we assume that a fraction
$\epsilon=0.8\%$ of $E_B$ is radiated away in $\tau_{\rm
  FRB}=1$ms. The efficiency $\epsilon$ we adopt is motivated
by~\cite{Lehner:2011aa}. The outgoing EM luminosity is estimated as
\begin{equation}
\label{eq:L_FRBs}
  L_{\rm FRB} \sim \epsilon\frac{E_{B}}{\tau_{\rm{\scalebox{0.8}{
          FRB}}}} \simeq 10^{42} \epsilon_{0.008}B_{12}^2 \tau_{\rm
    FRB,1}^{-1}\,\rm{erg\,s^{-1}}.
\end{equation}
The $L_{\rm FRB}$ estimate for each case we simulate is listed in
Table~\ref{table:results}. We also convert the luminosity to observed
flux density ($S_\nu$) at the detector in units of $\rm Jy$ using
$S_\nu=L_{\rm FRB}/4\pi D^2/\nu$, where $D$ is the luminosity distance to the
source, $\nu$ the radio telescope observing frequency. The flux density equation
can be written as
\begin{equation}
  \label{eq:L_FRBs}
  S_{\nu} \simeq 2.1\ {\rm Jy} \bigg(\frac{L_{\rm FRB}}{10^{41}\rm \ erg\,s^{-1}}\bigg) \bigg(\frac{D}{200\ \rm Mpc}\bigg)^{-2}\bigg(\frac{\nu}{1\ \rm GHz}\bigg)^{-1},
\end{equation}
where we chose a nominal observing frequency of 1 GHz (as is typical
of observed FRBs), and placed the source at the edge of the LIGO BNS
range. As shown in Table~\ref{table:results} the expected burst of the
EM radiation for a source at 200 Mpc has flux densities $\sim 0.1-30$
$\rm Jy$ and is fully consistent with observed FRB flux
densities~\cite{Mingarelli:2015bpo} that have been detected by current
radio telescopes such as CHIME, UTMOST, ASKAP, Parkes, and Arecibo.

We stress that the FRB in the model discussed here is not coming from
the collapse only. The inspiral magnetospheric interactions
contribute, making it possible to match the observed durations of
FRBs, the longest of which are challenging to match by the collapse
alone. The luminosity of the emission prior to
merger~\cite{Palenzuela:2013hu,Palenzuela:2013kra,Paschalidis:2013jsa,Ponce:2014sza,Ponce:2014hha}
is comparable to the post-collapse burst. We note while we were
writing our paper, the idea of an FRB from the prompt collapse alone
was also suggested in~\cite{Nathanail:2018jpu}.

\section{Conclusions}
\label{sec:conclusion}

In this paper, we performed a statistical study of the total mass and
mass ratio distribution of BNSs using the Galactic NS mass function
and population synthesis models in conjunction with recent constraints
on the nuclear EOS from GW170817. We find that up to $\sim 25\%$ of
all BNS mergers could result in prompt collapse. Moreover, our
analysis shows that most of the considered models of the BNS mass
function predict that the majority of prompt-collapse BNS mergers have
$q\gtrsim 0.8$, and that the larger $M_{\rm thres}$ is, the closer to
unity the $q$ distribution of prompt-collapse binaries
approaches. Prompt-collapse BNSs with $q>0.8$ are likely to unbind a
negligible amount of mass, and form negligibly small disks onto the
remnant BHs. Thus, neither detectable kilonovae nor sGRBs may
accompany the GWs from such prompt collapse BNSs. We referred to these
kilonovae- and sGRB-free BNS mergers as orphan. Our statistical study
suggests that most prompt-collapse BNS mergers may be
orphan. Therefore, the only remaining viable mechanism for powering
detectable electromagnetic counterparts from orphan BNS mergers is
related to magnetospheric effects.

We argued that the release of energy stored in the magnetosphere of
the merger remnant can match the {\it duration and power} of some FRBs
and that it also matches the luminosity of premerger magnetospheric
interactions. Thus, BNS mergers are promising sources of detectable,
non-repeating FRBs, as has been suggested before, and FRBs {\it may}
be the most promising electromagnetic counterpart of orphan BNS
mergers. The outgoing magnetospheric burst in these cases is rather
isotropic, making the detection of coincident FRB and GW signatures
possible. However, the most likely channel for such coincident
detections would be searches in LIGO data triggered by FRB detections,
because it is impossible for radio telescopes to follow up GW
detections on $\rm ms$ timescales.

We have also performed magnetohydrodynamic simulations in full general
relativity of different BNS configurations that undergo {\it prompt}
collapse. The stars are initially seeded with a dipolar magnetic field
that extends from the NS interior into the exterior. We computed the
energy stored in the magnetosphere following BH formation, and
estimated the outgoing electromagnetic luminosity produced. We find
luminosities $L_{\rm FRB}\sim 10^{40-42}B_{12}^2 \rm erg\,s^{-1}$,
which at the edge of the LIGO BNS range translate to flux densities of
0.1 to 30 $\rm Jy$, matching the flux density of previously observed
FRBs.


We close with a few caveats: First, our statistical analysis can be
refined as soon as ground based GW interferometers unveil the NS mass
function in BNSs; second if one is interested in the LIGO/Virgo {\it
  observed} mass function, the delay-time distribution should be
considered, which we do not account for here; third, some conclusions
in our work are based on the size of ejecta and BH disks found in
numerical relativity simulations of prompt-collapse BNS mergers. The
number of such simulations is small compared to simulations of BNS
mergers resulting in delayed collapse. Therefore, more high-resolution
simulations in full general relativity of BNSs resulting in prompt
collapse are necessary to solidify the results that such mergers
unbind negligible amounts of mass and form negligibly small disks onto
the remnant BH, and to find the ``critical'' mass ratio below which
appreciable mass ejection and disks occur. This critical mass ratio is
also likely to be equation-of-state dependent. Fourth, whether an FRB
signature from magnetospheric effects is luminous enough depends on
the NS surface magnetic field. We adopted a value of $\sim 10^{12}\,
\rm G$ , but FRB-level luminosities from magnetospheric interactions
are possible even from~$\sim 10^{11}\, \rm
G$~\cite{Palenzuela:2013hu}. Whether such regular pulsar magnetic
fields are present in these cases it is unclear, and this introduces a
source of uncertainty. If BNSs were to have only low magnetic fields,
this could make prompt-collapse BNS mergers completely orphan from an
electromagnetic point of view. However, on evolutionary grounds one of
the two components in a field BNS (i.e., not one that forms
dynamically in a cluster) is always anticipated to have a magnetic
field of $\sim 10^{11-12}\, \rm G$. This is because the NS that forms
second is not recycled, and hence its magnetic field is not ``buried''
during a recycling process, see, e.g.,~\cite{Lorimer:2008se} for a
review. In fact, the double pulsar J0737-3039 has provided a
spectacular confirmation of the evolutionary theory of double
NSs~\cite{Lorimer:2008se}. Pulsar B in J0737−3039 has an inferred
magnetic field of $1.6\times 10^{12}\, \rm
G$~\cite{Doublepulsar2004Sci...303}, while pulsar A is a millisecond
pulsar and has an inferred magnetic field of $6\times 10^9\, \rm
G$. In addition, the pulsar in the double NS J1906+0746 has an
inferred magnetic field strength $1.8\times 10^{12}\, \rm
G$~\cite{vanLeeuwen:2014sca}. Note that for stronger, near
magnetar-level magnetic fields a precursor burst of gamma-rays is
possible from BNS mergers~\cite{Metzger:2016mqu}. Another source of
uncertainty is that to explain FRBs we need to know the efficiency for
converting the total magnetospheric power output to radio waves. If
this efficiency is less than 1\%, then it is not likely that FRBs can
be accounted for by magnetospheric effects. Finally, with our code we
are able to obtain only crude estimates of the energy in the
magnetosphere. A more accurate assessment of the full FRB signature in
the model considered here requires a code (such as that
of~\cite{Palenzuela:2013hu,Palenzuela:2013kra}) that can evolve
through inspiral, merger and prompt collapse to magnetosphere release,
while smoothly matching the ideal magnetohydrodynamic stellar interior
to a force-free exterior. Such a simulation is currently lacking and
will be the subject of future work of ours.\\


\acknowledgements We thank Richard O'Shaughnessy for pointing us to
the Synthetic Universe website. We also thank S. L. Shapiro and
R. Thompson for useful discussions. This work has been supported in
part by National Science Foundation (NSF) Grant Grant PHY-1912619 at
the University of Arizona, by NSF Grants PHY-1602536 and PHY-1662211,
and NASA Grant 80NSSC17K0070 at the University of Illinois at
Urbana-Champaign. Simulations were in part run on the Perseus cluster
at Princeton University. This work made use of the Extreme Science and
Engineering Discovery Environment (XSEDE) under grant numbers
TG-PHY180036 and TG-PHY190020.

\bibliographystyle{apsrev4-1}        
\bibliography{references}            
\end{document}